# The Simulation and Mapping of Building Performance Indicators based on European Weather Stations

A.W.M. (Jos) van Schijndel

*Department of the Built Environment, Eindhoven University of Technology, Netherlands*

## ABSTRACT

Due to the climate change debate, a lot of research and maps of external climate parameters are available. However, maps of indoor climate performance parameters are still lacking. This paper presents a methodology for obtaining maps of performances of similar buildings that are virtually spread over whole Europe. The produced maps are useful for analyzing regional climate influence on building performance indicators such as energy use and indoor climate. This is shown using the Bestest building as a reference benchmark. An important application of the mapping tool is the visualization of potential building measures over the EU. Also the performances of single building components can be simulated and mapped. It is concluded that the presented method is efficient as it takes less than 15 minutes to simulate and produce the maps on a 2.6GHz/4GB computer. Moreover, the approach is applicable for any type of building.

## 1. Introduction

Buildings and their systems can be considered as complex dynamic systems. The control and operation of building systems are of eminent importance regarding durability, health and costs. Energy use can be drastically reduced if building systems are properly controlled and optimal operated. At the same time comfort can be improved and costs can significantly be lowered. The goal of Computational Building Physics is to search for physics based models behind the empirical facts in order to control the built environment. Furthermore Computational Building Physics studies the built environment on several scales (see Figure 1)

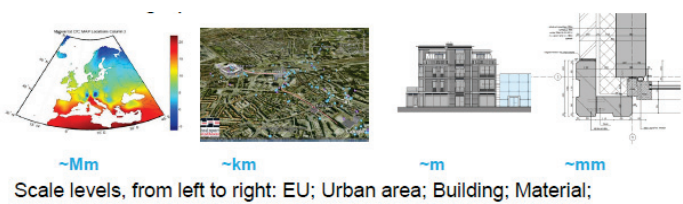

Figure 1. Scale levels involved within building physics research

Building physics research applications are so far almost always limited to the scale levels of material, building and urban area. In this paper we present a building physics research application of the scale level of the EU.

### 1.1 Related work

In this Section we will focus on two important building related research areas where EU mappings are already common techniques. First, we start with cultural heritage and climate change. Grossi et al. (2007) are using maps to visualize the prediction of the evolution in frost patterns due to climate change during the 21$^{st}$ century and the potential damage to historic structures and archeological remains in Europe. Figure 2 shows an exemplarily result of the application of a freezing event map.

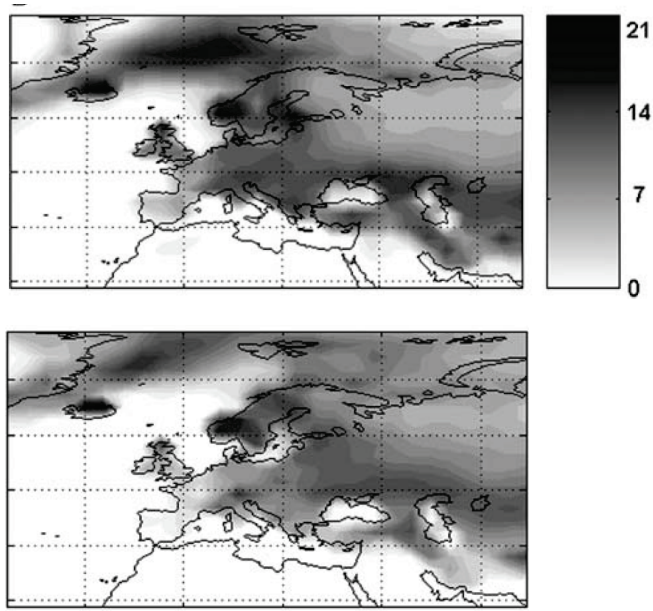

Figure 2. Pan-European maps of average yearly freezing events in 30 years period 1961–1990 (top) and far future 2070–2099 (bottom) by Grossi et al. (2007).

Similar maps as presented in Figure 2 are used to show the expected reduction of freezing and lowering the potential for frost shattering of porous building stone. The underlying data for these maps are based on regional climate models. This is the second research area where EU maps are commonly used. There is an enormous amount of literature on climate change and mapping. Therefore we illustrate the use of these maps by one state of the art regional climate model: REMO (Jacob 1997, Larsen 2010). Figure 3 shows the twenty-five-year mean modeled wind at 10 m height over Central Europe using REMO with a 10 km resolution.

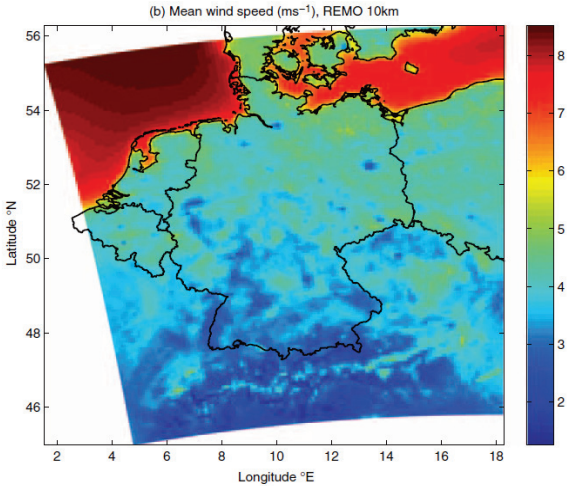

Figure 3. Twenty-five-year mean modeled wind at 10 m height over the entire domain REMO 10 km resolution (Larsen 2010)

Maps like figure 3 are suited for wind energy assessment application in Northern Europe. Moreover, literature of the related work shows that a lot of EU maps of external climate parameters are available.

### 1.2 Goal and Outline

The maps presented in the previous section are all based on external climate parameters. However, the goal of this work is produce maps that include also *indoor* climate performance parameters.

The outline of the paper is as follows: Section 2 presents the methodology of the application for obtaining maps of performances of similar buildings that are virtual spread over whole Europe. The produced maps are useful for analyzing regional climate influence on building performance indicators such as energy use and indoor climate. Section 3 shows a benchmark of the EU mapping of the Bestest building. Section 4 comprehends a case study on the simulation and mapping of the effect of improved glazing using the Bestest building as a reference.

## 2. Methodology

The methodology used for obtaining the required simulation results and maps can be divided into three steps. These are presented in the following sections.

### 2.1 External climate files

Over 130 external hourly-based climate files were produced using commercially available software (Meteonorm 2011) using so-called wac format. Figure 4 presents the distribution of the locations over Europe.

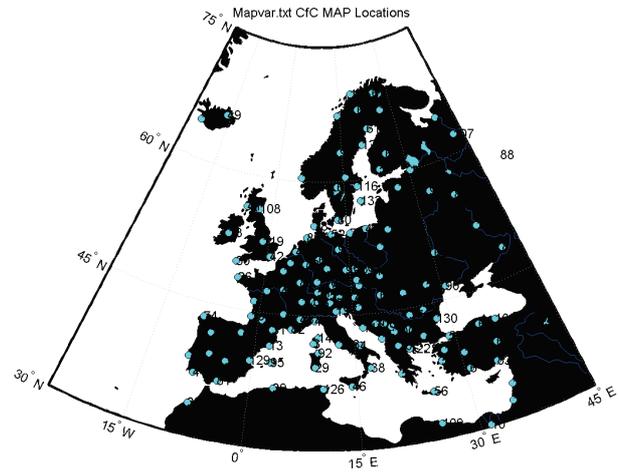

Figure 4. The distributions of the locations of the external climates in Europe.

Each climate file includes hourly based values for the common used external climate parameters: Horizontal global solar radiation [W/m$^2$] (ISGH), Diffuse solar radiation [W/m$^2$] (ISD), Cloud cover [0-1] (CI), Air temperature [$^o$C] (TA), Relative humidity [%] (HREL), Wind speed [m/s] (WS), Wind direction [0-360$^o$] (WD), Rain intensity [mm/h] (RN), Long wave radiation [W/m$^2$] (ILAH).

### 2.2 Whole building simulation model

The whole building model originates from the thermal indoor climate model ELAN which was already published in 1987 (de Wit et al. 1988). Separately a model for simulating the indoor air humidity (AHUM) was developed. In 1992 the two models were combined and programmed in the MatLab environment. Since that time, the model has constantly been improved using newest techniques provided by recent MatLab versions. The current hourly-based model HAMBase, is part of the Heat, Air and Moisture Laboratory (HAMLab 2011), and is capable of simulating the indoor temperature, the indoor air humidity and energy use for heating and cooling of a multi-zone building. The physics of this model is extensively described by de Wit (2006). The main modeling considerations are summarized below. The HAMBase model uses an integrated sphere approach. It reduces the radiant temperatures to only one node. This has the advantage that complicated geometries can easily be modeled. Figure 5 shows the thermal network.

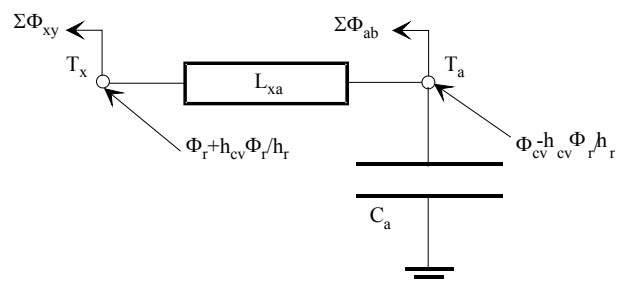

Figure 5. The room radiative model as a thermal network

Where $T_a$ is the air temperature and $T_x$ is a combination of air and radiant temperature. $T_x$ is needed to calculate transmission heat losses with a combined surface coefficient. $h_r$ and $h_{cv}$ are the surface weighted mean surface heat transfer coefficients for convection and radiation. $\Phi_r$ and $\Phi_{cv}$ are respectively the radiant and convective part of the total heat input consisting of heating or cooling, casual gains and solar gains.

For each heat source a convection factor can be given between 0 and 1 by the user. For example, for air heating this factor is close to 1 and for radiative heat sources a factor ranging from 0.4 – 0.6 can be used. The factor for solar radiation depends on the window system and the amount of radiation falling on furniture. $C_a$ is the heat capacity of the air. $L_{xa}$ is a coupling coefficient (de Wit et al. 1988):

$$L_{xa} = A_t h_{cv} \left(1 + \frac{h_{cv}}{h_r}\right) \quad (1)$$

$\sum \Phi_{ab}$ is the heat loss by air entering the zone with an air temperature $T_b$. $A_t$ is the total area. In case of ventilation $T_b$ is the outdoor air temperature. $\sum \Phi_{xy}$ is transmission heat loss through the envelope part y. For external envelope parts $T_y$ is the sol-air temperature for the particular construction including the effect of atmospheric radiation.

The admittance for a particular frequency can be represented by a network of a thermal resistance ($1/L_x$) and capacitance ($C_x$) because the phase shift of $Y_x$ can never be larger than $\pi/2$. To cover the relevant set of frequencies (period 1 to 24 hours) two parallel branches of such a network are used giving the correct admittance's for cyclic variations with a period of 24 hours and of 1 hour. This means that the heat flow $\Phi_{xx}(tot)$ is modeled with a second order differential equation. For air from outside the room with temperature $T_b$ a loss coefficient $L_v$ is introduced. This model is summarized in Figure 6.

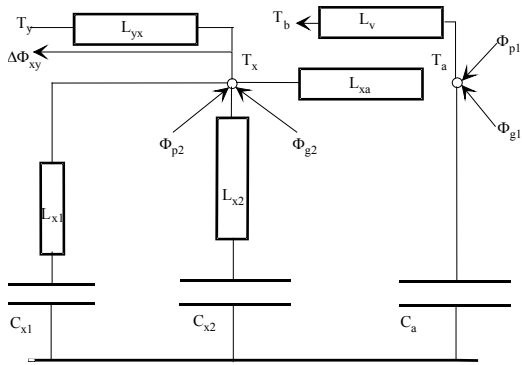

Figure 6. The thermal ventilation model for one zone

In a similar way a model for the air humidity is made. Only vapour transport is modeled, the hygroscopic curve is linearized between RH 20% and 80%. The vapour permeability is assumed to be constant. The main differences are: a) there is only one room node (the vapour pressure) and b) the moisture storage in walls and furniture, carpets etc is dependent on the relative humidity and temperature.

The HAMbase model is adapted in such a way that all climate (.wac) files in a directory are automatically processed. For each climate file a corresponding output file is produced containing hourly based values for the indoor climate and heating and cooling power.

A separate Matlab mfile is developed for calculating annual means and peak values for each location (i.e. wac file) and together with the longitude and latitude stored in a single file suitable for mapping purposes.

### 2.3 Mapping of the results

A MatLab mfile was developed for the visualization of the just mentioned mapping file. For the exact details of this mfile, we refer to the HAMLab website (HAMLab 2012).

## 3. Benchmark: EU Mapping of the Bestest building

### 3.1 Bestest using HAMBase

The Bestest (ASHRAE, (2001)) is a structured approach to evaluate the performance of building performance simulation tools. The evaluation is performed by comparing results of the tested tool relative to results by reference tools. The procedure requires simulating a number of predefined and hierarchal ordered cases. Firstly, a set of qualification cases have to be modeled and simulated. If the tool passes all qualification cases the tool is considered to perform Bestest compliant. In case of compliance failure the procedure suggests considering diagnostic cases to isolate its cause. Diagnostic cases are directly associated with the qualification cases (Judkoff and Neymark 1995). The first qualification case, case 600 (see Figure 7) was used for the performance comparison.

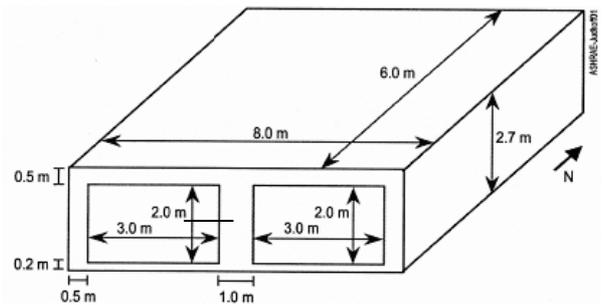

Figure 7. Bestest case 600 geometry

The thermal part of HAMBase has been subjected to a standard method of test (Bestest ASHRAE, (2001)), with satisfactory results. The accompanying climate file of the Bestest is based on weather station near Denver (USA). For further details, see Table I.

Table I Comparison of the HAMBase model with some cases of the standard test.

| Case | Nr. Simulation of | model | test min | ..max |
|---|---|---|---|---|
| 600ff | mean indoor temperature [°C] | 25.1 | 24.2 .. | 25.9 |
| 600ff | min. indoor temperature [°C] | -17.9 | -18.8 .. | -15.6 |
| 600ff | max. indoor temperature [°C] | 64.0 | 64.9 .. | 69.5 |
| 900ff | mean indoor temperature [°C] | 25.1 | 24.5.. | 25.9 |
| 900ff | min. indoor temperature [°C] | -5.1 | -6.4.. | -1.6 |
| 900ff | max. indoor temperature [°C] | 43.5 | 41.8.. | 44.8 |
| 600 | annual sensible heating [MWh] | 4.9 | 4.3.. | 5.7 |
| 600 | annual sensible cooling [MWh] | 6.5 | 6.1.. | 8.0 |
| 600 | peak heating [kW] | 4.0 | 3.4.. | 4.4 |
| 600 | peak sensible cooling [kW] | 5.9 | 6.0.. | 6.6 |
| 900 | annual sensible heating [MWh] | 1.7 | 1.2.. | 2.0 |
| 900 | annual sensible cooling [MWh] | 2.6 | 2.1 .. | 3.4 |
| 900 | peak heating [kW] | 3.5 | 2.9 .. | 3.9 |

### 3.2 EU mapping

Now we apply the method of Section 2 in order to produce EU maps of certain (performance) parameters.

#### 3.2.1 External climate

We start with the outdoor climate. For these maps simulation is not required because the information is already present in the climate (.wac) files. Figure 8 shows the expected north-south temperature distribution. The influence of higher altitude weather stations near the Alps is also visible.

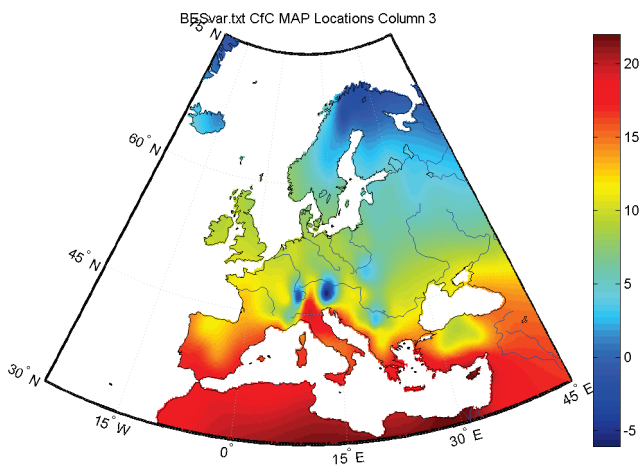

Figure 8. Mean annual external air temperature [°C] (This figure relies on color, see digital version of the paper)

The relative humidity distribution is shown in Figure 9.

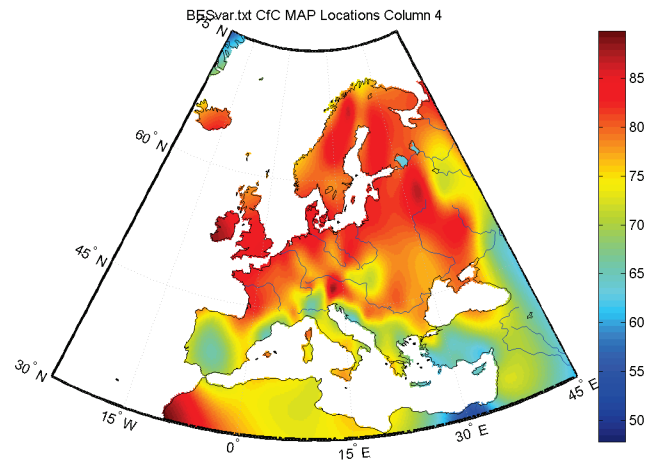

Figure 9. Mean annual external relative humidity [%] (This figure relies on color, see digital version of the paper)

Quite remarkably there is a peak in relative humidity at the east part of the Alps corresponding with the low temperature (see Figure 8) which is not present at the west side of the Alps.

#### 3.2.2 Indoor Climate

The simulation of all 130 weather stations of the Bestest case 600 building takes less than 15 minutes using HAMBase on a 2.6GHz/4GB computer. Figure 10 shows the mean annual indoor air temperature distribution.

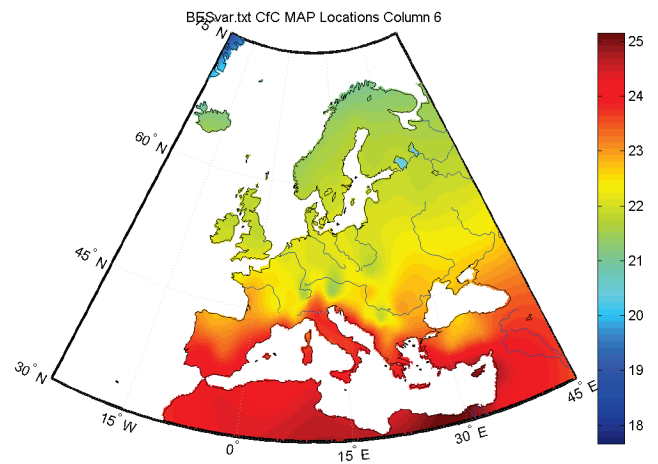

Figure 10. Mean annual indoor air temperature [°C] of the Bestest case 600 building (This figure relies on color, see digital version of the paper)

As expected the mean annual indoor temperature is within the 20 to 27 °C range (the upper left part of the map near Greenland is not accurate due to the absence of weather stations). Although it is not within the Bestest, it could be interesting to simulate the mean annual indoor relative humidity (in case that there are no internal moisture gains). This is presented in Figure 11.

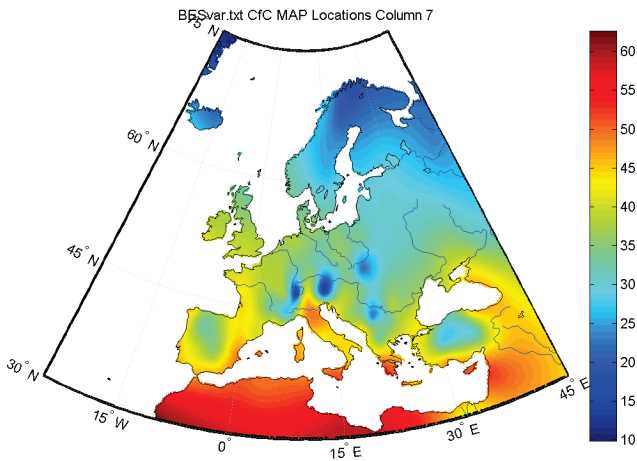

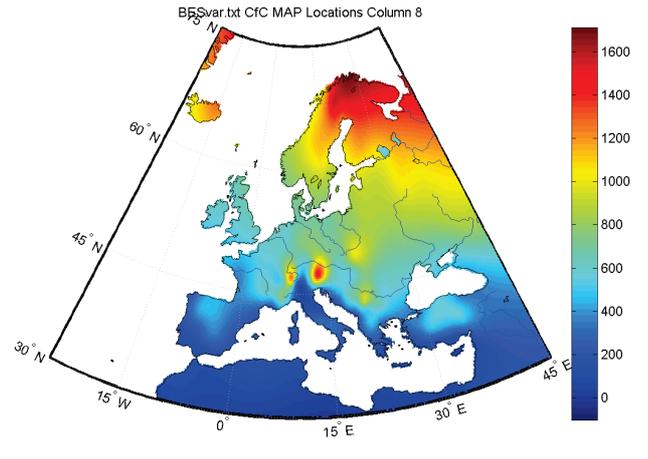

Figure 11. Mean annual indoor relative humidity [%] of the Bestest case 600 building (This figure relies on color, see digital version of the paper)

Figure 13. Mean annual heating power [W] of the Bestest case 600 building (This figure relies on color, see digital version of the paper)

As expected at the cold regions (see Figure 8) the indoor relative humidity is low.

Because the Bestest case 600 building has 12 m$^2$ of window area facing south it is interesting to visualize the mean annual solar gain [W] entering the room. This is shown in Figure 12.

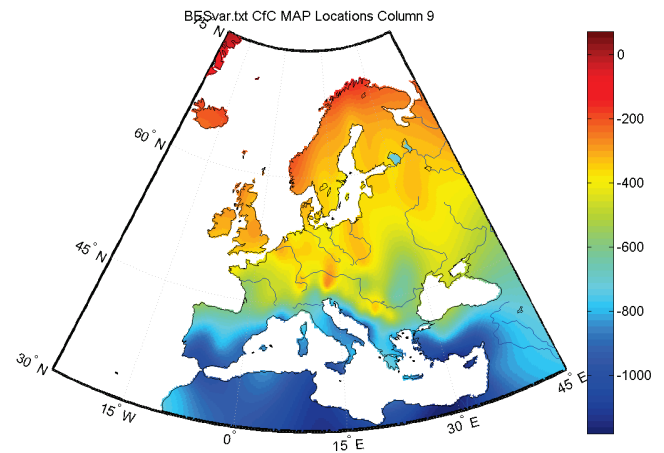

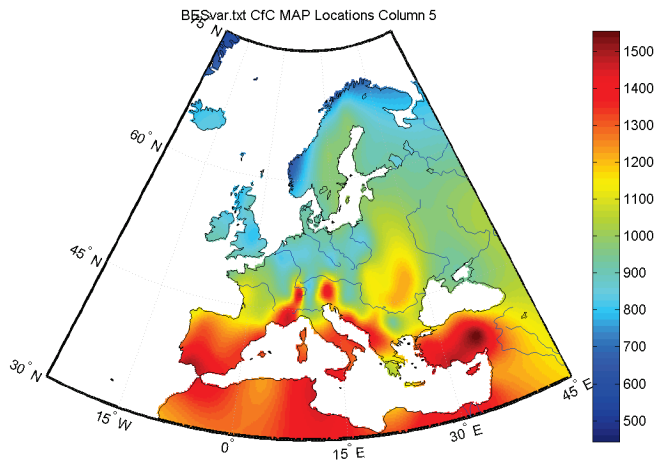

Figure 14. Mean annual cooling power [W] of the Bestest case 600 building (see also digital version of the paper for color)

Figure 12. Mean annual solar gain [W] of the Bestest case 600 building (This figure relies on color, see digital version of the paper)

Figure 12 is in line with what we expect.

### 3.2.3 Heating & Cooling

The energy use for heating and cooling the Bestest case 600 building is shown in figure 13 through 16. The mean annual heating and cooling power maps are quite as expected. Also the peak heating power (Figure 15) has an intuitive map. Remarkably, the peak cooling power map (Figure 16) shows a quite uniform distribution ranging from 5.5-7 kW.

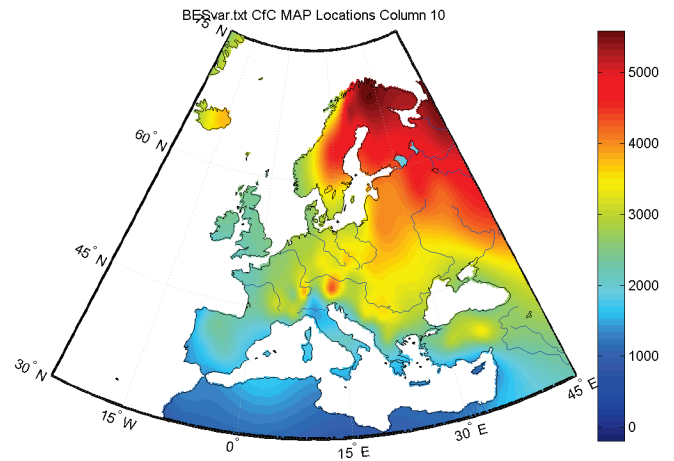

Figure 15. Peak heating power [W] the of Bestest case 600 building (see also digital version of the paper for color)

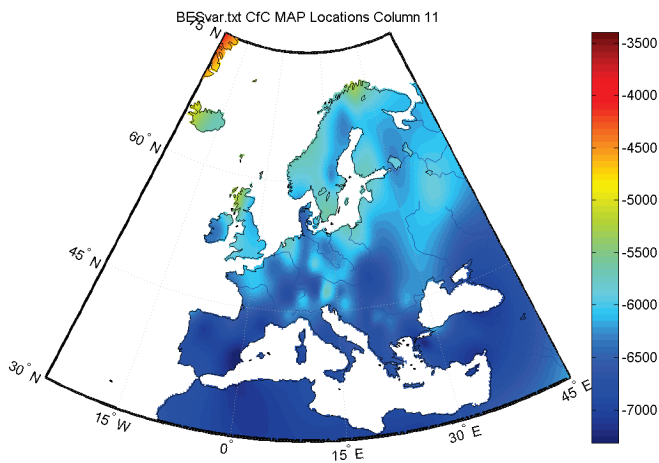

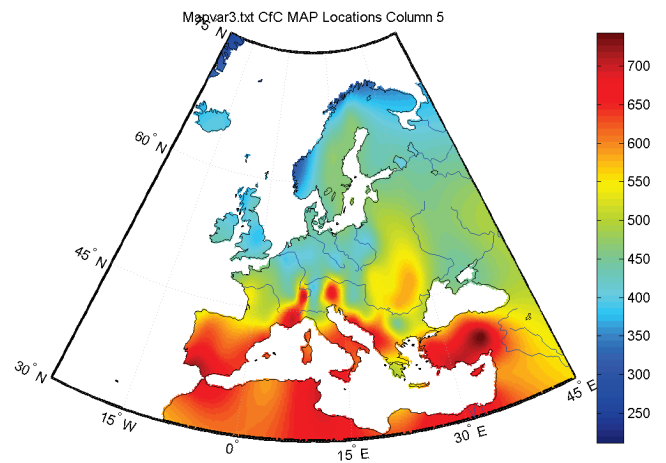

Figure 16. Peak cooling power [W] of the Bestest case 600 building (see also digital version of the paper for color)

Figure 17. Reduction of mean annual solar gain [W] of the Bestest case 600 building due to the improved glazing (see also digital version of the paper for color)

## 4. EU Mapping of single building components effects

In the previous Section the main performances of a reference building model subjected to EU weather stations data, were visualized using EU maps. This leads to a new interesting application. If we change a single building component we can visualize the single effect of this change by subtracting these (new) results with the *reference* case. In order to demonstrate this idea we study the effect of replacing the original glazing (U value = 3 W/m$^2$K; Solar gain factor = 0.787) of the case 600 building with improved glazing (U value = 1.5 W/m$^2$K; Solar gain factor = 0.40) called *variant*.

The method to obtain the results below was as follows: (1) The building was simulated and mapped again using the new glazing parameters. (2) A net **change** effect map was obtained by subtracting two maps, i.e. *reference - variant* map. (3) A **percentage change** effect map was obtained by using the formula: (1- *variant/reference*) * 100% map.

### 4.1 Indoor climate change

The results show only minor changes in the mean indoor climates of respectively 1 $^o$C and 1%. Because the solar gain factor is reduced from 0.787 to 0.40 we expect substantial differences between the reference and the variant. Figure 17 shows the result:

### 4.2 Heating & Cooling

#### 4.2.1 Mean annual heating reduction

Figures 18 through 20 show the three types of maps for the mean annual heating. Figure 18 provides the mean annual heating of the building with improved glazing:

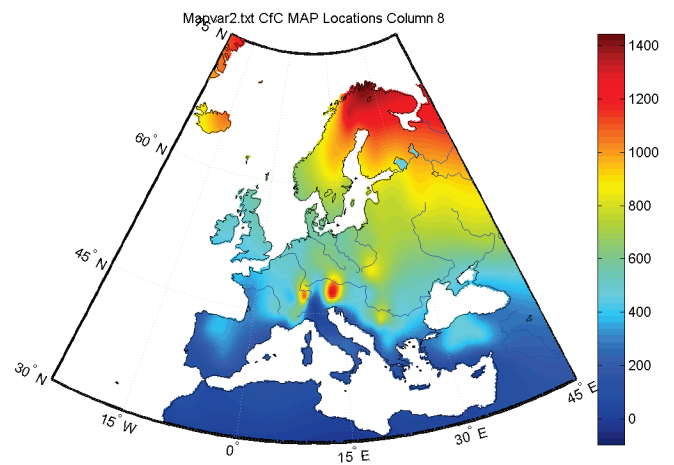

Figure 18. Mean annual heating power [W] of the Bestest case 600 building with improved glazing (see also digital version of the paper for color)

Using the map of mean annual heating power of the reference building (i.e. Figure 13), a net change effect map was obtained by subtracting two maps, i.e. reference map minus variant map. This is shown in figure 19.

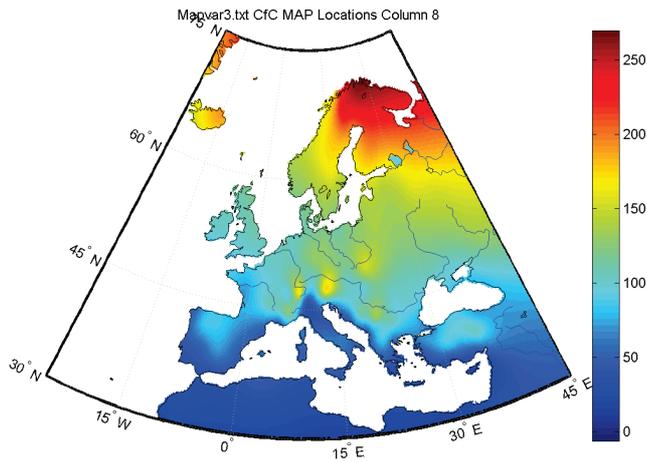

Figure 19. Change in mean annual heating power [W] of the Bestest case 600 building with improved glazing (see also digital version of the paper for color)

The highest reduction in heating amount can be observed in the Nordic countries (up to 250Wy x 8.76 kh/y = 2190 kWh). The lowest reduction can be found in the Mediterranean as expected. The reduction can also be expressed as a percentage change effect map, using the formula: (1-variant/reference) * 100%. This is presented in figure 20.

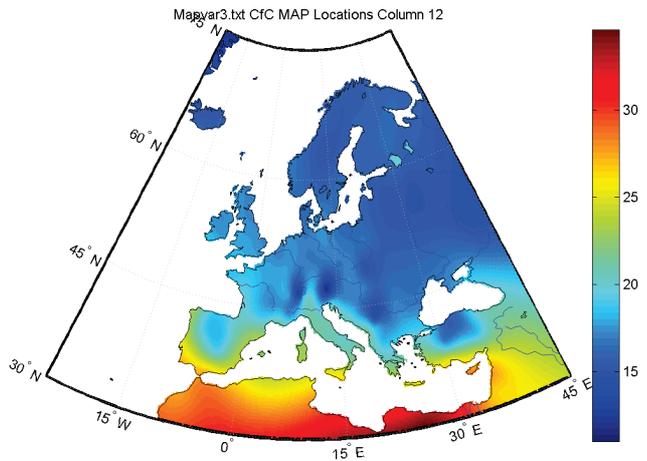

Figure 20. Percent change in mean annual heating power [%] of the Bestest case 600 building with improved glazing (see also digital version of the paper for color)

Please note that although the Mediterranean has a relatively high percent change, the absolute values for the mean annual heating power are low, so almost no heating energy savings are expected in this region. Secondly in areas where improved glazing do have impact, the relative improvement is quite uniform between 10 and 20%.

### 4.2.2  Mean annual cooling reduction

Figure 21 provides the mean annual cooling reduction.

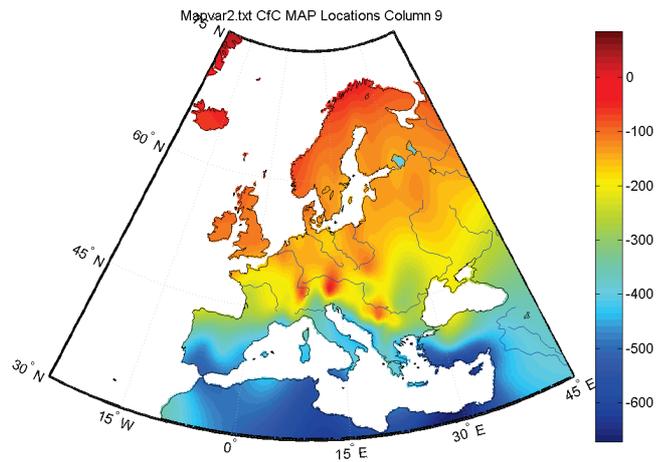

Figure 21. Change in mean cooling power [W] of the Bestest case 600 building with improved glazing (see also digital version of the paper for color)

The highest reduction in cooling amount can be observed in the Mediterranean (up to -500Wy x 8.76 kh/y = 4400 kWh). The reduction can also be expressed as a percentage change effect map. This is presented in figure 22.

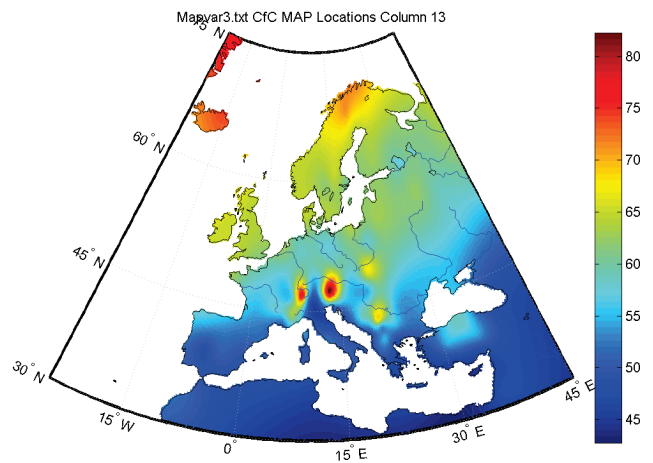

Figure 22. Percent change in mean annual cooling power [%] of the Bestest case 600 building with improved glazing (This figure relies on color, see digital version of the paper)

Although the percent change at the Alps is relatively high, the absolute values for the mean annual cooling power are low, so almost no cooling energy savings are expected in this region.

### 4.2.3  Peak heating reduction

Figure 23 shows the relative the peak heating power reduction.

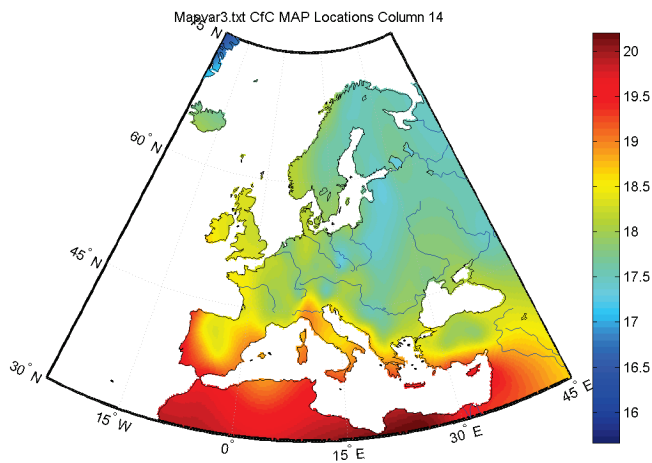

Figure 23. Reduction of the peak heating power [%] of the Bestest case 600 building (see also digital version of the paper for color)

Remarkably, figure 23 shows a quite uniform relative change in peak heating power reduction.

*4.2.4 Peak cooling reduction*

Figure 24 shows the absolute peak cooling power reduction.

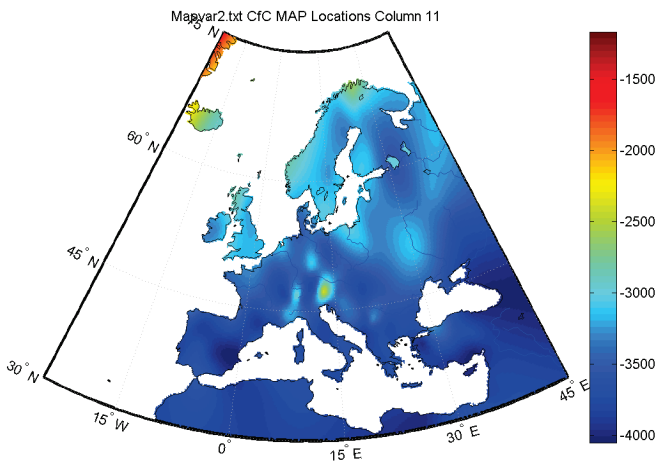

Figure 24. Reduction of the peak cooling power [W] of the Bestest case 600 building (see also digital version of the paper for color)

Figure 24 shows a uniform relative change in cooling power reduction (3 – 4 kW).

## 5. Conclusions

The produced maps are useful for analyzing regional climate influence on building performance indicators such as energy use and indoor climate. This is shown using the Bestest building as a reference benchmark. An important application of the mapping tool is the visualization of potential building measures over the EU. Also the performances of single building components can be simulated and mapped. It is concluded that the presented method efficient as it takes less than 15 minutes to simulate and produce the maps on a 2.6GHz/4GB computer. Moreover, the approach is applicable for any type of building

*Future research*

Within the mentioned EU FP7 project 'Climate for Culture', detailed EU external climate files are currently under development for the period 1960 – 2100 using the REMO model (Jacob et al. 1997) and a moderate climate scenario. With these future external climate files we will be able to predict future building performance indicators. Together with the EU mapping tool this could be helpful to locate EU regions with the highest impact on the specific building performances.

## 6. Acknowledgement

This work was inspired and sponsored by the EU FP7 project 'Climate for Culture'.